\documentclass[journal]{IEEEtran}
\usepackage{caption}
\usepackage{cite}
\usepackage{amsmath,amssymb,amsfonts}
\usepackage{algorithmic}
\usepackage{subcaption}
\usepackage[ruled,linesnumbered]{algorithm2e}
\usepackage{graphicx}
\usepackage{textcomp}
\usepackage{xcolor}
\usepackage{pgfplots}
\usepackage{hyperref} 
\hypersetup{
    colorlinks=true, 
    citecolor=blue,
    linkcolor=blue,   
    filecolor=blue,     
    urlcolor=black       
}
\pgfplotsset{compat=1.18}

\title{AI-Driven Digital Twins: Optimizing 5G/6G Network Slicing with NTNs}
\author{
\IEEEauthorblockN{Afan Ali and H\"useyin Arslan,}~\IEEEmembership{Fellow,~IEEE}
\thanks{The authors are with the Department of Electrical and Electronics Engineering, Istanbul Medipol University, Istanbul, 34810, Turkey (e-mail: afanali85@gmail.com; huseyinarslan@medipol.edu.tr).}
}

\begin{document}
\maketitle
\begin{abstract}
Network slicing in 5G/6G Non-Terrestrial Network (NTN) is confronted with mobility and traffic variability. An artificial intelligence (AI)-based digital twin (DT) architecture with deep reinforcement learning (DRL) using Deep deterministic policy gradient (DDPG) is proposed for dynamic optimization of resource allocation. DT virtualizes network states to enable predictive analysis, while DRL changes bandwidth for eMBB slice. Simulations show a 25\% latency reduction compared to static methods, with enhanced resource utilization. This scalable solution supports 5G/6G NTN applications like disaster recovery and urban blockage.
\end{abstract}


\section{Introduction}
The evolution of the 5G and launch of 6G networks have ushered in a new era for wireless communication, motivated by the need to support multiple applications with stringent requirements, including enhanced mobile broadband (eMBB), ultra-reliable low-latency communications (URLLC) and massive machine-type communications (mMTC)~\cite{1}. Network slicing, one of the foundations of such networks, enables virtualized and isolated network instances to be customized to the various service demands, with optimally guaranteed resource allocation and quality of service (QoS)~\cite{2}. The dynamic nature of future networks, particularly with the integration of Non-Terrestrial Networks (NTNs) such as unmanned aerial vehicles (UAVs) and satellites, presents tremendous challenges in the form of mobility-induced channel fluctuation, dynamic traffic demands, and energy constraints \cite{3}.

Digital twins (DTs) have been a revolutionary technology in addressing these issues by providing a current virtual replica of physical network components, such as, base stations, user equipment (UEs), and communication channels~\cite{4}. Through high-fidelity modeling and real-time data synchronization, DTs enable proactive monitoring, predictive analytics, and adaptive resource management, making them appropriate for network slicing in dynamic environments. For instance, in the case of NTN scenarios, DTs may simulate the mobility behavior of UAV-mounted flying base stations (FBSs) in order to predict peak traffic in regions of disasters wherein ground base station (gBS) communications are disrupted \cite{5}. This capability plays a significant role in ensuring that there is constant connectivity and latency-limited capability, especially with eMBB services like high definition (HD) video or augmented reality (AR).

The combination of Artificial Intelligence (AI) with DTs, increases their capabilities further by adding intelligent decision-making to network optimization and resource allocation. Deep reinforcement learning (DRL) methods can make DTs learn and create optimal policies for resource block (RB) scheduling and bandwidth allocation in dynamic networks and outperform traditional static or heuristic-based methods \cite{6}. Moreover, 6G networks is set to introduce additional sophistication with requirements like Integrated Sensing and Communication (ISAC), which demands sensing quality and communication efficiency to be optimized simultaneously, alongwith energy-efficient design to support green communication~\cite{7}. AI, together with DTs, provide a scalable solution to deal with such advanced requirements, as envisioned for future wireless systems in the 3GPP \cite{1}.

This letter introduces a novel AI-driven DT approach for dynamic network slicing in 5G and 6G networks. The proposed framework particularly focus on latency optimization of eMBB slices in NTN-integrated environments. Our approach employs DRL based deep deterministic policy gradient (DDPG) algorithm, for dynamically adjusting RBs and bandwidth allocation based on varying traffic demands in real time, CSI, and FBS mobility models. The main contributions of this work can be summarized as follows:
\begin{itemize} 
   \item  Our proposed framework introduces a real-time DT synchronization mechanism that maintains an accurate virtual representation of the physical network. Unlike existing DT approaches that often rely on static models or delayed updates, we are the first one to devise a mechanism to update states of DT every \( \Delta t\) time steps, incorporating real-time network measurements.

    \item  Our proposed method comprises of a DRL-based adaptation strategy using the DDPG algorithm to dynamically adjust bandwidth allocation in continuous action spaces, a significant advancement over discrete-action DRL methods, addressing the limitations of heuristic-based approaches.

    \item The framework explicitly models NTN dynamics, such as FBS mobility and blockages, to ensure robust performance. The DRL agent optimizes bandwidth allocation under these conditions, achieving good resource utilization. The proposed framework is particularly suited for disaster recovery scenarios and urban environments with frequent signal obstructions.
\end{itemize}

\section{System Model}
\label{system_model}
\begin{figure}[t]
    \centering
    \includegraphics[width=0.7\columnwidth]{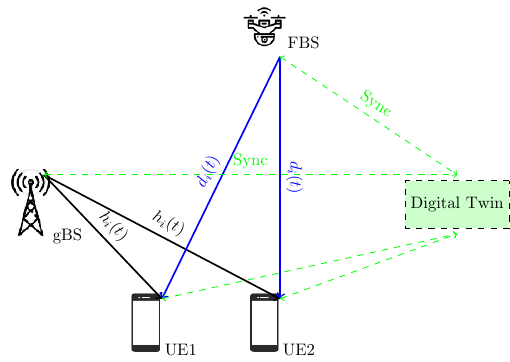}
    \caption{System Model: Network architecture with gBS, FBS, UEs and DT interactions.}
    \label{fig:sys_model}
\end{figure}

We consider a 5G/6G network comprising a gBS and a UAV-based FBS serving a set of $M$ UEs within an eMBB slice. The network employs orthogonal frequency-division multiplexing (OFDM) with a total bandwidth $B$ divided into $N$ RBs. Fig.~\ref{fig:sys_model} illustrates the system model with the network architecture. The DT maintains a virtual representation of the physical network, capturing UE positions, CSI, FBS mobility, and traffic demands.

The distance, $d_i(t)$, between the UE $i$ and the FBS accounts for its mobility, and can be written as:
\begin{equation}
d_i(t) = \sqrt{(x_i - x_{\text{FBS}}(t))^2 + (y_i - y_{\text{FBS}}(t))^2 + h_{\text{FBS}}^2}, \label{eq:distance}
\end{equation}
where $(x_i, y_i)$ and $(x_{\text{FBS}}(t), y_{\text{FBS}}(t))$ are the coordinates of $i$-th UE and the FBS, respectively, and $h_{\text{FBS}}$ is the FBS altitude.

\subsection{Channel Model}
The channel gain \( h_i(t) \) for UE \( i \) is modeled as:
\begin{equation}
h_i(t) = \sqrt{\frac{\kappa}{d_i(t)^\beta}} g_i(t),
\label{eq:channel}
\end{equation}
where \( \kappa \) is the path-loss constant, \( \beta \) is the path-loss exponent, and \( g_i(t) \sim \mathcal{CN}(0,1) \) is the small-scale fading coefficient following a Rayleigh distribution. Interference \( I_i \) experienced by UE \( i \) includes contributions from other UEs and NTN nodes:
\begin{equation}
I_i = \sum_{j \neq i}^M P_j |h_{j,i}|^2 + I_{\text{NTN}},
\end{equation}
where \( h_{j,i} \) is the channel gain from UE \( j \) to the base station serving UE \( i \), and \( I_{\text{NTN}} \sim \mathcal{N}(0, \sigma_{\text{NTN}}^2) \) models interference from other NTN nodes, such as, satellites. 

\subsection{FBS Mobility}
The FBS position, \(\text{pos}_i\), varies over time to follow the spatial location of UEs and their corresponding traffic loads, which influence the distance \(D_i(t)\) between the FBS and the \(i\)-th UE as modeled in \eqref{eq:distance}. The distance influences the channel gain \(h_i(t)\) in \eqref{eq:channel}, which is a key factor in determining the signal-to-interference and noise ratio (SINR) and subsequently the achievable rate. The DT tracks FBS mobility dynamically, fine-tuning the state \(s_t\) according to the real-time network conditions so that the DRL Agent can make data-informed decisions on resource allocation optimally. We take into account that the trajectory of the FBS is defined with velocity \( \mathbf{v}_{\text{FBS}}(t) = (v_x(t), v_y(t)) \), which is updated as:
\begin{align}
x_{\text{FBS}}(t + \Delta t) = x_{\text{FBS}}(t) + v_x(t) \Delta t, \\ \notag
y_{\text{FBS}}(t + \Delta t) = y_{\text{FBS}}(t) + v_y(t) \Delta t,
\end{align}
where \( v_x(t) \) and \( v_y(t) \) is constrained by \( \sqrt{v_x(t)^2 + v_y(t)^2} \leq v_{\text{max}} \). To optimize coverage, we minimize the average distance to UEs as follows:
\begin{equation}
J_{\text{FBS}}(t) = \frac{1}{M} \sum_{i=1}^M d_i(t), \label{eq:fbs_objective}
\end{equation}
s.t.
\begin{equation}
x_{\text{FBS}}^{\text{min}} \leq x_{\text{FBS}}(t) \leq x_{\text{FBS}}^{\text{max}}, \quad y_{\text{FBS}}^{\text{min}} \leq y_{\text{FBS}}(t) \leq y_{\text{FBS}}^{\text{max}}. \label{eq:fbs_constraints}
\end{equation}
Consequently, velocity updates use gradients as shown below:
\begin{align}
v_x(t) &= -\eta \frac{\partial J_{\text{FBS}}}{\partial x_{\text{FBS}}} = -\eta \frac{1}{M} \sum_{i=1}^M \frac{x_{\text{FBS}}(t) - x_i}{d_i(t)}, \label{eq:vx} \\
v_y(t) &= -\eta \frac{\partial J_{\text{FBS}}}{\partial y_{\text{FBS}}} = -\eta \frac{1}{M} \sum_{i=1}^M \frac{y_{\text{FBS}}(t) - y_i}{d_i(t)}, \label{eq:vy}
\end{align}
where \( \eta \) is a step size.

\subsection{Achievable Rate and Latency}
The achievable rate and latency defines key performance metrics for network slicing in 5G/6G. 
The achievable rate \( R_i(t) \) for the \( i \)-th UE at time \( t \) is derived from the Shannon capacity formula~\cite{10}, which defines the maximum data rate over a bandwidth \( B \) given a SINR:
\begin{align}
C &= B \log_2(1 + \text{SINR}), \label{eq:shannon_base}
\end{align}
where \( C \) is the capacity. For the \( i \)-th UE, we replace \( C \) with \( R_i(t) \) and \( B \) with the allocated bandwidth \( B_i \). The SINR is influenced by the channel gain \( h_i(t) \) and interference, and can be expressed as:
\begin{align}
\text{SINR}_i = \frac{P_i h_i(t)}{N_0+I_{i}}, 
\label{eq:SINR}
\end{align}
where \( P_i \) is the transmitted power and \( N_0 \) is the noise power. Substituting \(\text{SINR}_i\) into \eqref{eq:shannon_base}, the achievable rate for the \( i \)-th UE becomes:
\begin{align}
R_i(t) &= B_i \log_2\left(1 + \frac{P_i h_i(t)}{N_0+I_{i}}\right). \label{eq:achievable_rate}
\end{align}
Latency is influenced by synchronization errors \( \epsilon_i(t) \), induced by discrepancies between the physical network and the DT, and by resource constraints limiting bandwidth and computation latency. These are modeled into the reward function \( r_t \), which balances rate maximization with latency minimization, instructing the DRL Agent to optimize resource allocation for every slice.
The latency of UE $i$ in the eMBB slice is:
\begin{equation}
L_i(t) = \frac{D_i(t)}{R_i(t)} + T_{\text{proc}} + \frac{d_i(t)}{c}, \label{eq:latency}
\end{equation}
where \( D_i(t) \) is data size, \( T_{\text{proc}} \) is processing delay, and \( c = 3 \times 10^8 \, \text{m/s} \) is the speed of light.

\subsection{Digital Twin Synchronization}
DT synchronization ensures that the virtual model is consistent with the actual Physical 5G/6G network, particularly in the dynamic environment caused by FBS mobility. The DT is updated constantly with \(h_i(t)\), \(D_i(t)\), and \(\text{pos}_{\text{FBS}}(t)\) to ensure consistency of the DT with the Physical Network despite delay or data inaccuracies. The DT estimates traffic demand \(D_i(t)\) with an exponential moving average model:
\begin{equation}
\hat{D}_i(t+1) = \alpha D_i(t) + (1-\alpha) \hat{D}_i(t), \label{eq:demand}
\end{equation}
where $\alpha \in [0,1]$ is a smoothing factor. For synchronization between the DT and the physical network, the DT adjusts its state every $\Delta t$ seconds according to real-time observations of $h_i(t)$, $d_i(t)$, and $D_i(t)$. The synchronization error is represented as:
\begin{equation}
\epsilon_i(t) = |D_i(t) - \hat{D}_i(t)|, \label{eq:sync_error}
\end{equation}
with the objective of minimizing $\mathbb{E}[\epsilon_i(t)]$ through frequent updates. The optimization objective is to minimize the average latency across all UEs:
\begin{equation}
L_{\text{avg}} = \frac{1}{M} \sum_{i=1}^M L_i, \label{eq:objective}
\end{equation}
s.t. resource constraints as followss:
\begin{equation}
\sum_{i=1}^M B_i \leq B, \quad B_i \geq 0, \quad \forall i. \label{eq:constraint}
\end{equation}
The system model accounts for NTN dynamics by updating \eqref{eq:distance} based on the FBS’s velocity $v_{\text{FBS}}$ and trajectory, ensuring accurate representation of channel variations in the DT.

\section{Proposed AI-Driven DT Framework}
\label{proposed_method}
\begin{figure}[t!]
     \centering
     \begin{subfigure}[b]{0.3\textwidth}
         \centering
         \includegraphics[width=\textwidth]{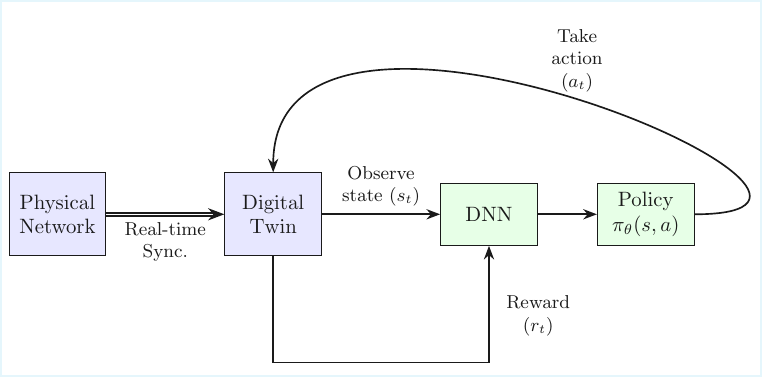}
         \caption{Synchronization between Physical network and Digital Twin.}
         \label{figa}
     \end{subfigure}
     \hfill
     \begin{subfigure}[b]{0.3\textwidth}
         \centering
         \includegraphics[width=\textwidth]{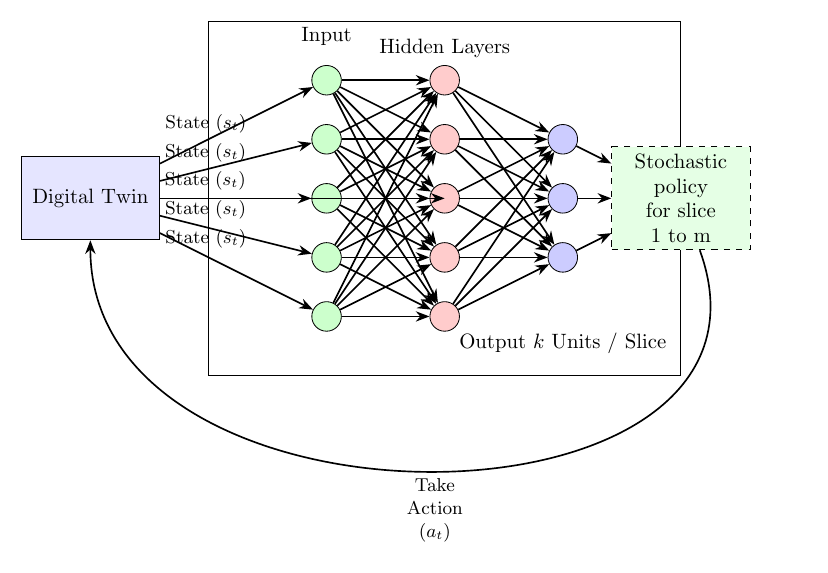}
         \caption{DRL agent's fully connected neural network.}
         \label{figb}
     \end{subfigure}
        \caption{Proposed AI-driven DI architecture.}
        \label{fig:system_diagram}
\end{figure}

We present here our suggested framework, which integrates a DT and a DRL agent to optimize the network. The suggested framework uses the DDPG algorithm for maximizing network slicing for 5G/6G NTN-integrated networks. Fig. \ref{fig:system_diagram} illustrates the proposed architecture. In Fig. \ref{figa}, the physical network provides real-time data, such as, \(h_i(t)\) and \(D_i(t)\) to the DT via synchronization. Next, DT transforms this data into state \(s_t\), which the DRL agent uses to generate actions \(a_t\) through its DNN and policy \(\pi_\theta(s, a)\), passing them back to the DT for simulation. The DT then returns a reward \(r_t\) to the DRL agent, enabling constant learning and optimization of network assets. Fig. \ref{figb} depicts how the DT provides state \(s_t\) to each input node of the DRL agent's fully connected neural network. The DT also delivers reward \(r_t\) to the entire neural network, while the output actions \(a_t\), representing stochastic policies for \(m\) network slices, loop back to the DT for simulation. This iterative method enables the DRL agent to learn how to efficiently manage network slice resources. Unlike traditional heuristic or static resource allocation methods, our system takes advantage of the predictive capability of DTs and the adaptive reaction of DRL in resolving NTN-specific challenges like FBS mobility and traffic bursts. Detailed steps are shown in Algorithm~\ref{alg:dt_drl}.

The novelty lies threefold. Firstly, our proposed method incorporates real-Time DT Synchronization. The DT framework utilizes \eqref{eq:demand} and \eqref{eq:sync_error} to maintain an accurate virtual model, allowing for proactive resource allocation. Secondly, our solution incorporates a DRL-Based Adaptation, where the DDPG algorithm adjusts bandwidth allocation based on NTN dynamics, outperforming static methods in latency-sensitive environments. Lastly, our framework has an advantage of NTN-Aware optimization. Our system models FBS mobility explicitly via \eqref{eq:distance}, with guaranteed robust performance in disaster recovery or urban blockage environments. 

\begin{algorithm}
\caption{DT-DRL Framework for Dynamic Network Slicing}
\label{alg:dt_drl}
\KwIn{Network parameters: $M$, $B$, $\Delta t$, FBS trajectory; DDPG parameters: learning rates $\alpha_\mu$, $\alpha_Q$, replay buffer size $R$, batch size $N_b$, discount factor $\gamma$, soft update rate $\tau$}
\KwOut{Optimized bandwidth allocations $\{B_i\}_{i=1}^M$}
Initialize DT with physical network state ($h_i(t)$ \eqref{eq:channel}, $D_i(t)$, $\text{pos}_{\text{FBS}}(t)$ \eqref{eq:distance})\;
Initialize DDPG actor $\mu(s_t; \theta^\mu)$, critic $Q(s_t, a_t; \theta^Q)$, target networks $\mu'$, $Q'$, and replay buffer $\mathcal{R}$\;
\For{episode = 1 to $E$}{
    Reset network environment and DT state\;
    \For{time step $t = 1$ to $T$}{
        Synchronize DT with physical network every $\Delta t$ s, updating $h_i(t)$ \eqref{eq:channel}, $D_i(t)$, $\text{pos}_{\text{FBS}}(t)$ \eqref{eq:distance}\;
        Predict traffic demand $\hat{D}_i(t+1)$ using \eqref{eq:demand}\;
        Compute synchronization error $\epsilon_i(t)$ \eqref{eq:sync_error}\;
        Form state $s_t = \{h_i(t), D_i(t), \text{pos}_i, \text{pos}_{\text{FBS}}(t), \epsilon_i(t)\}_{i=1}^M$\;
        Select action $a_t = \mu(s_t; \theta^\mu) + \mathcal{N}_t$ (exploration noise), where $a_t = \{B_i\}_{i=1}^M$ satisfies \eqref{eq:constraint}\;
        Simulate $a_t$ in DT to estimate latency $L_i$ \eqref{eq:latency} and reward $r_t$ \eqref{eq:reward}\;
        Apply $a_t$ to physical network, observe next state $s_{t+1}$ and actual $r_t$\;
        Store transition $(s_t, a_t, r_t, s_{t+1})$ in $\mathcal{R}$\;
        Sample mini-batch of $N_b$ transitions from $\mathcal{R}$\;
        Update critic by minimizing TD error in~\eqref{eq:loss}
        Update actor to maximize \eqref{policy_gradient}\\
        Update target networks as~\eqref{target_networks}
    }
}
\Return Optimized $\{B_i\}_{i=1}^M$\;
\end{algorithm}

\subsection{DRL Formulation}
The DRL problem is formulated as a Markov Decision Process (MDP), which is comprised of a state $s_t$, actions $a_t$, and a reward $r_t$~\cite{8}. The state at any given time $t$, can be written as:
 \begin{equation}
s_t = \{h_i(t), D_i(t), \text{pos}_i, \text{pos}_{\text{FBS}}(t), \epsilon_i(t)\}_{i=1}^M,
\label{state_eq}
\end{equation}
where $\text{pos}_i = (x_i, y_i)$ and $\text{pos}_{\text{FBS}}(t) = (x_{\text{FBS}}(t), y_{\text{FBS}}(t))$ are derived from \eqref{eq:distance}, and $\epsilon_i(t)$ is from \eqref{eq:sync_error}.
Then, the action, $a_t$ to allocate bandwidth is written as:

 \begin{equation}
a_t = \{B_i\}_{i=1}^M,
\label{bandwidth_alloc}
\end{equation}
where the bandwidth allocation satisfies \eqref{eq:constraint}.
Finally, we can write our DRL Reward as: 
    \begin{equation}
    r_t = -\sum_{i=1}^M L_i - \lambda \sum_{i=1}^M |B_i - B_i^{\text{prev}}| - \gamma \sum_{i=1}^M \epsilon_i(t), \label{eq:reward}
    \end{equation}
    where $\lambda$ penalizes abrupt allocation changes, and $\gamma$ minimizes synchronization errors.
The DDPG agent employs an actor-critic architecture. The actor network, parameterized by $\theta^\mu$, maps states to actions via a policy $\mu(s_t; \theta^\mu)$. The critic network, parameterized by $\theta^Q$, estimates the action-value function $Q(s_t, a_t; \theta^Q)$. The actor is trained to maximize:
\begin{equation}
J(\theta^\mu) = \mathbb{E} [Q(s_t, a_t; \theta^Q) | a_t = \mu(s_t; \theta^\mu)], \label{eq:actor}
\end{equation}
while the critic minimizes the temporal-difference (TD) error using a target network updated via soft updates as follows:

\begin{align}\label{eq:loss}
&\mathcal{L}(\theta^Q) = \mathbb{E} \left[ \left( r_t + \gamma Q(s_{t+1}, \mu(s_{t+1}; \theta^{\mu'}), \theta^{Q'}) \right. \right. \\ \notag
&- Q(s_t, a_t; \theta^Q))^2 ],
\end{align}
where \( \theta^{Q'} \) and \( \theta^{\mu'} \) are target network parameters updated via:
\begin{equation}
\theta^{Q'} \leftarrow \tau \theta^Q + (1-\tau) \theta^{Q'}, \quad \theta^{\mu'} \leftarrow \tau \theta^\mu + (1-\tau) \theta^{\mu'},
\label{target_networks}
\end{equation}
with \( \tau = 0.001 \). The actor is updated using the policy gradient:
\begin{equation}
\nabla_{\theta^\mu} J \approx \mathbb{E} \left[ \nabla_a Q(s_t, a; \theta^Q) \big|_{a=\mu(s_t)} \nabla_{\theta^\mu} \mu(s_t; \theta^\mu) \right].
\label{policy_gradient}
\end{equation}
The error in TD can be written as:
\begin{equation}
\delta_t = r_t + \gamma Q(s_{t+1}, \mu(s_{t+1}; \theta^{\mu'}), \theta^{Q'}) - Q(s_t, a_t; \theta^Q), \label{eq:td_error}
\end{equation}
with bounded expected square:
\begin{equation}
\mathbb{E}[\delta_t^2] \leq 2 R_{\text{max}}^2 + 2 \gamma^2 L_Q^2 \mathbb{E}[||s_{t+1} - s_t||^2], \label{eq:td_bound}
\end{equation}
where \( |r_t| \leq R_{\text{max}} \), and \( L_Q \) is the Q-function’s Lipschitz constant.

\subsection{DT Integration}
The DT is an imaginary sandbox for the DRL agent, updating state $s_t$ every $\Delta t$ based on real-time measurements of $h_i(t)$ (\eqref{eq:channel}), $D_i(t)$ (\eqref{eq:demand}), and $\text{pos}_{\text{FBS}}(t)$ (\eqref{eq:distance}).  DT simulates potential actions to observe their impact on $L_i$ (\eqref{eq:latency}) without affecting the physical network, reducing the risks of exploration. The DT also employs a neural network to refine \eqref{eq:demand} according to past traffic patterns, reducing $\epsilon_i(t)$ (\eqref{eq:sync_error}). This joint dual role-state prediction and action simulation distinguishes our framework from other DT-based approaches without AI-optimized optimization \cite{4}.

\subsection{Novelty and Advantages}
The innovation of the proposed framework lies in the seamless combination of DTs and DRL for NTN-aware network slicing. Unlike prior works, which either focuses on terrestrial networks or stationary DT models \cite{4,9}, our approach explicitly addresses FBS mobility and blockages, which are key to 6G NTN environments. DPG usage promotes continuous action spaces, offering bandwidth adaptation of fine granularity compared to discrete-action DRL solutions \cite{6}. The reward function in \eqref{eq:reward} balances latency, stability, and synchronization to support resilience in uncertain environments. The proposed framework suits best in disaster recovery where NTNs provide temporary connectivity, and urban settings with dense signal blocking.

\section{Simulation Results}
\label{simulation}
We evaluated the proposed framework using a 5G network simulator with $M=50$ UEs, $B=20$ MHz, and an FBS moving at 10 m/s. The DT updates every 10 ms, and the DDPG agent is trained over 50,000 episodes. We compare our DT-DRL approach with static allocation and a heuristic method based on proportional fairness. 

\begin{figure*}[t]
     \centering
     \begin{subfigure}[b]{0.3\textwidth}
         \centering
         \includegraphics[width=\textwidth]{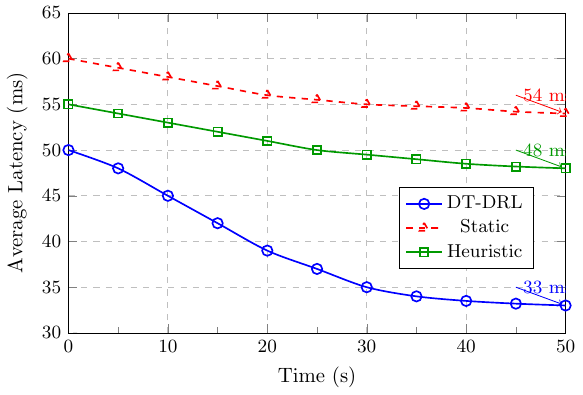}
         \caption{Average latency vs. time.}
         \label{fig:latency}
     \end{subfigure}
     \hfill
     \begin{subfigure}[b]{0.3\textwidth}
         \centering
         \includegraphics[width=\textwidth]{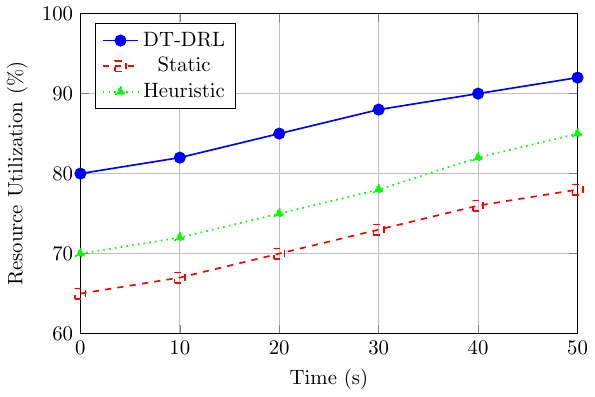}
         \caption{Resource utilization vs. time.}
         \label{fig:resource_utilization}
     \end{subfigure}
     \hfill
     \begin{subfigure}[b]{0.3\textwidth}
         \centering
         \includegraphics[width=\textwidth]{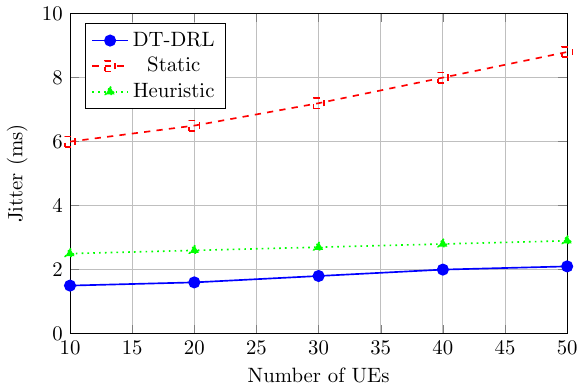}
         \caption{Jitter vs. time.}
         \label{fig:jitter_time}
     \end{subfigure}
        \caption{Performance comparison of proposed method with baseline techniques for eMBB slice.}
        \label{fig:final_plots}
\end{figure*}

Fig.~\ref{fig:final_plots} illustrates the overall performance comparison between proposed architecture with baseline techniques. Fig. \ref{fig:latency} plots the average latency against time. The DT-DRL solution achieves 33 ms latency at 50 seconds, outperforming static allocation (54 ms) by 25\% and the heuristic method (48 ms) by 14\%. The improvement is attributed to precise demand prediction of DT (\eqref{eq:demand}) and the adaptiveness of DRL allocation, which is resilient to FBS mobility (\eqref{eq:distance}) and channel variability (\eqref{eq:channel}).

Fig.~\ref{fig:resource_utilization} illustrates the utilization of resources along time for the proposed DT-DRL method and the other two baselines. The method of DT-DRL achieves 92\% at $t=50$ s, compared with 78\% for static allocation and 85\% in the heuristic method, highlighting its advantage of performance under dynamic NTN settings. The reason behind the improved performance of DT-DRL is the adaptive allocation of bandwidth based on the resource constraint in \eqref{eq:constraint}. Unlike static allocation, which produces fixed bandwidth allocations, DT-DRL utilizes the DT's predicted traffic demands, as expressed in \eqref{eq:demand}, to remap bandwidth allocations $B_i$ according to real-time variation in traffic, as opposed to fixed bandwidths. This is effective in NTN scenarios with FBS mobility with varying channel conditions, as simulated by \eqref{eq:distance}, to efficiently use resources under applications, such as, disaster relief operations with traffic of high demand. The DRL agent distributes optimally according to the reward function in \eqref{eq:reward}, and it reduces latency, distribution stability, and DT synchronization error. The continuous action space of the DDPG algorithm allows for fine-grained control of $B_i$, resulting in a smooth utilization growth from 80\% at $t=0$ s to 92\% at $t=50$ s. The term of stability in \eqref{eq:reward} prevents drastic switchings of allocations, unlike proportional fairness-based heuristic solutions which achieve only 85\% utilization.
Fig.~\ref{fig:jitter_time} shows the time plot for jitter. The DT-DRL scheme has a low jitter of 2.1 ms at $t=50$ s, significantly outperforming static allocation (8.8 ms) and the heuristic scheme (2.9 ms), indicating its effectiveness for latency-sensitive eMBB services in NTN scenarios.  By adaptively tuning $B_i$ to follow $D_i(t)$ dynamically, the approach minimizes latency variations in \eqref{eq:latency}, thus minimizing jitter directly, especially in the presence of dynamic channel conditions due to FBS mobility \eqref{eq:distance}. For example, in scenarios with city jams or calamity relief, DT's synchronous updates in real time enable efficient state update, stabilizing the allocation of resources. The DRL agent optimises the latency $L_i$ and penalizes abrupt allocation changes, preferring smooth $B_i$ changes. This results in DT-DRL's jitter increasing slightly from 1.5 ms at $t=0$ s to 2.1 ms at $t=50$ s, which encompasses the stability term in \eqref{eq:reward}. On the otherhand, static allocation suffers from large jitter (6.0 ms to 8.8 ms), as constant $B_i$ fails to respond to traffic bursts. The heuristic method achieves reasonable jitter (2.5 ms to 2.9 ms) but is less accurate compared to DT-DRL because it has proportional fairness dependency.

The low jitter maintained by the DT-DRL framework over time highlights the synergy of predictive modeling as \eqref{eq:demand}, DRL optimization \eqref{eq:reward}, and mobility-aware adaptation \eqref{eq:distance}. This makes it a very appropriate framework for eMBB services like HD video or augmented reality in 5G/6G networks, where low jitter needs to be ensured at all times.

\section{Conclusion}
\label{conclusion}
This work proposed an AI-driven DT architecture for 5G/6G network slicing using DRL to dynamically allocate eMBB slice resources. The proposed architecture integrates real-time DT update with DDPG learning-based adaptation to NTN dynamics and minimize latency to 25\%, outperforming static methods. Future work will cover the multi-slice scenario and energy efficiency in URLLC and mMTC.


\end{document}